\documentclass[12pt]{iopart}

\usepackage{graphicx}
\begin{document}

\title[Thickness dependence of structural and transport properties of Co-doped BaFe$_2$As$_2$]{Thickness dependence of structural and transport properties of Co-doped BaFe$_2$As$_2$ on Fe buffered MgO substrates}

\author{Kazumasa Iida, Jens H\"{a}nisch, Sascha Trommler, Silvia Haindl\\Fritz Kurth, Ruben H\"{u}hne, Ludwig Schultz\\and Bernhard Holzapfel}

\address{Leibniz-Institut f\"{u}r Festk\"{o}rper-und Werkstoffforschung (IFW) Dresden\\P.\,O.\,Box 270116, 01171 Dresden, Germany}
\ead{k.iida@ifw-dresden.de}
\begin{abstract}
We have investigated the influence of the superconducting layer thickness, $d$, on the structural and transport properties of Co-doped BaFe$_2$As$_2$ films deposited on Fe-buffered MgO substrates by pulsed laser deposition. The superconducting transition temperature and the texture quality of Co-doped BaFe$_2$As$_2$ films improve with increasing $d$ due to a gradual relief of the tensile strain. For $d\geq90\,\rm nm$ an additional 110 textured component of Co-doped BaFe$_2$As$_2$ was observed, which leads to an upward shift in the angular dependent critical current density at $H$\,$\parallel$\,${c}$. These results indicate that the grain boundaries created by the 110 textured component may contribute to the $c$-axis pinning.
\end{abstract}

\submitto{\SUST}

\maketitle

\section{Introduction}
Crystalline quality, structural parameters and surface morphology of thin films are generally influenced by their thickness due to several factors such as strain, defect formation and change in growth mode. Such changes affect the physical properties of thin films. Hence, whenever new functional thin films are prepared, it always becomes immediately interesting to explore the effect of layer thickness on the structural and physical properties.

Among the family of newly discovered Fe-based superconducting compounds, increasing the layer thickness of FeSe and FeSe$_{0.5}$Te$_{0.5}$ thin films improves the superconducting transition temperature, $T_{\rm c}$, presumably due to lattice distortion by strain\,\cite{01,02}. 

For Co-doped BaFe$_2$As$_2$ (Ba-122) films, only thickness studies regarding the buffer layers have been reported to date. Tarantini $et$ $al$ and Lee $et$ $al$ have found that single crystalline (La,Sr)(Al,Ta)O$_3$ substrates  with 100 unit cells of epitaxial SrTiO$_3$ resulted in the highest $T_{\rm c}$ and the largest critical current density, $J_{\rm c}$\,\cite{03,04}. Even higher $T_{\rm c}$ and sharper out-of-plane and in-plane textures of the Fe/Ba-122 bilayers can be realized for 20\,nm thick epitaxial Fe buffer layers\,\cite{05}. However, no investigation of the effect of layer thickness of Co-doped Ba-122 on the structural and transport properties have been published to date. In this article, we report on the influence of layer thickness on the structural and transport properties of the Fe/Ba-122 bilayer system with a fixed Fe layer thickness.

\section{Experiment}
Epitaxial, smooth Fe buffer layers (20\,nm) were prepared by a two-step process, which involves a room temperature deposition of Fe on MgO (001) single crystalline substrates by pulsed laser deposition, PLD, followed by a high-temperature annealing at 750\,$^\circ$C, both in a UHV chamber (base pressure of 10$^{-10}$\,mbar). Prior to the Fe deposition, the substrate was heated to 1000\,$^\circ$C, held at this temperature for 30\,min, subsequently cooled to room temperature for cleaning. A KrF excimer laser (248\,nm) has been employed at a frequency of 5\,Hz for the deposition with an energy density of 3--5\,Jcm$^{-2}$ on the target. After the Fe buffer preparation, Co-doped Ba-122 layers were deposited at 750\,$^\circ$C with a laser repetition rate of 10\,Hz. Each deposition step was monitored by reflection high-energy electron diffraction, RHEED. The layer thickness, $d$, was varied in the range of 30\,nm to 225\,nm by controlling the number of laser pulses. Each layer thickness was confirmed by cross-sectional focused ion beam, FIB, cuts on multiple sample areas. The nominal composition of the PLD target was Ba:Fe:Co:As\,=\,1:1.84:0.16:2. The detailed target preparation can be found in reference\,\cite{06}. The phase purity of the target was determined by x-ray diffraction using a standard Bragg-Brentano geometry with Co-K$\alpha$ radiation. All the observed peaks were identified as Co-doped Ba-122. The lattice parameters refined via Rietveld analyses were $a=0.39586(2)$\,nm and $c=1.29825(6)$\,nm, respectively. 

Surface morphology of the films was observed by atomic force microscopy, AFM. Out-of-plane texture and phase purity were investigated by x-ray diffraction in Bragg-Brentano geometry with Co-K$\alpha$ radiation. In-plane orientation of both Fe and Co-doped Ba-122 were investigated by using the 110 and 103 poles respectively in a texture goniometer operating with Cu-K$\alpha$ radiation. In order to evaluate the in-plane and out-of-plane lattice parameters of Co-doped Ba-122 precisely, high resolution reciprocal space maps, RSM, around the 109, 10\underline{11} and 11\underline{10} reflections were performed with Cu-K$\alpha$ radiation. Here, the 204 reflection of MgO was used as a reference to eliminate any errors by a misalignment of the substrate. 

After the structural characterization, Au layers were deposited on the films by PLD at room temperature followed by ion beam etching to form bridges of 0.5\,mm width and 1\,mm length for transport measurements.  Superconducting properties were measured in a Physical Property Measurement System (PPMS, Quantum Design) by a standard four-probe method with a criterion of 1\,$\rm\mu Vcm^{-1}$ for evaluating $J_{\rm c}$. In the angular-dependent $J_{\rm c}$ measurements, $J_{\rm c}(\Theta)$, the magnetic field, $H$, was applied in the maximum Lorentz force configuration ($H$ perpendicular to $J$) at an angle $\Theta$ measured from the $c$-axis. $T_{\rm c}$ is defined as 50\% of the normal state resistance at 30\,K.  

\section{Results and discussion}
\begin{figure}[t]
	\centering
		\includegraphics[width=8.5cm]{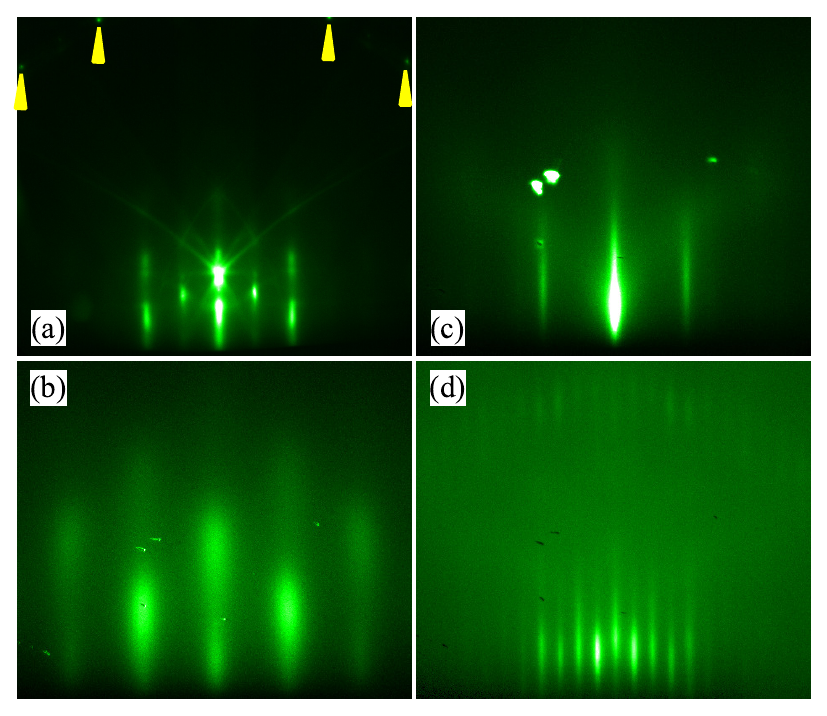}
		\caption{Representative RHEED images of Fe/Ba-122 bilayer ($d$=150\,nm). (a) MgO single crystalline substrate at room temperature after the heat treatment. Yellow arrows indicate the diffraction spots lie on the Laue circles. (b) Fe at room temperature at the end of deposition, (c) Fe at 750\,$^\circ$C, and (d) Co-doped Ba-122 at room temperature. The incident electron beam is along the MgO [110] azimuth.} 
\label{fig:figure1}
\end{figure}

The diffraction pattern in the RHEED images of MgO substrate shows a series of spots lying on the Laue circles, indicative of a perfectly flat surface (figure\,\ref{fig:figure1}(a)). For the Fe buffer preparation, the RHEED images of Fe in figure\,\ref{fig:figure1}(b) confirm the epitaxial growth even at room temperature for $d$=150\,nm. The diffraction spots turn into streaks with increasing temperature (figure\,\ref{fig:figure1}(c)), indicative of smoothing of the surface\,\cite{07}. For Co-doped Ba-122, the diffraction patterns with long streaks centered at positions on the Laue circles is typical for a multilevel surface (figure\,\ref{fig:figure1}(d)); i.e. for a high number of smooth terraces separated by steps\,\cite{07}. Additionally, the spacing of the observed streaks indicates a surface reconstruction, which is consistent with the observation on single crystals reported in reference\,\cite{08}.

\begin{figure}
	\centering
		\includegraphics[width=8.5cm]{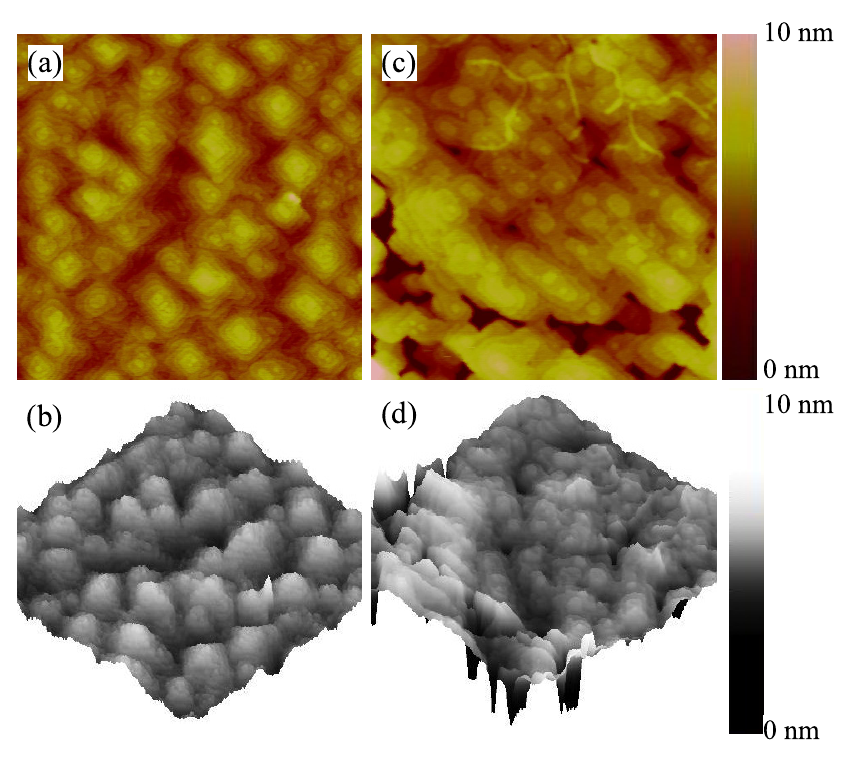}
		\caption{(a) 2-Dimensional and (b) 3-dimensional AFM images (1$\mu$m$\times$1$\mu$m) of Fe/Ba-122 bilayer ($d$=150\,nm) exhibits a large number of terraced islands. (c), (d) The corresponding images of the 30\,nm thick film show that the grains are smaller and not well connected.} 
\label{fig:figure2}
\end{figure}

The AFM image of this film in figures\,\ref{fig:figure2}\,(a) and (b) further confirmed that the surface was flat with a root mean-square roughness, $R_{\rm rms}$, of 0.83\,nm. The AFM image also shows that Co-doped Ba-122 grows in the terraced-island mode with an average step height of 0.65\,nm, which is almost identical to half the lattice parameter $c$. All the films in this study have the same surface morphology, and their surface roughness are summarized in Table\,\ref{tab:table1}. However, the grains of the 30\,nm thick film are smaller and not well connected compared with the other films (figures.\,\ref{fig:figure2}(c) and (d)).       

\begin{figure}
	\centering
		\includegraphics[width=8.5cm]{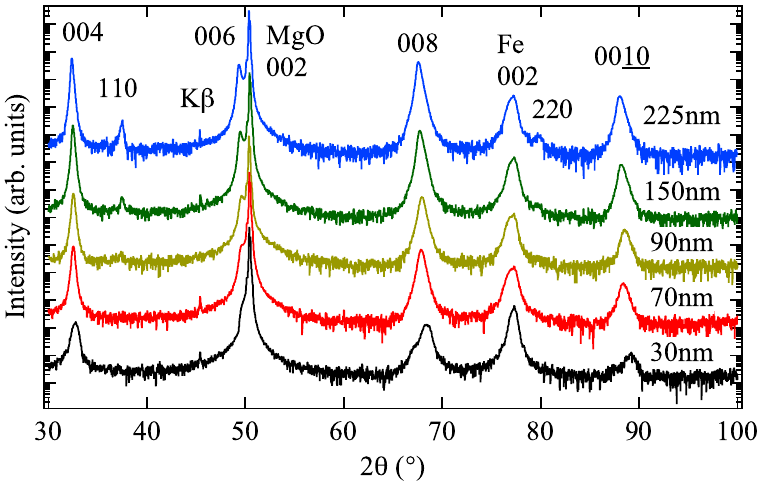}
		\caption{The $\theta\rm/2\theta$-\,scans of Fe/Ba-122 bilayers with various layer thicknesses on (001) MgO substrates. Intensity of the 110 reflection is observed to increase with $d$.} 
\label{fig:figure3}
\end{figure}

The $\theta\rm/2\theta$-\,scans for the Fe/Ba-122 bilayers with different layer thickness do not show any secondary phases (figure\,\ref{fig:figure3}). The pronounced 00$l$ reflections of Co-doped Ba-122 together with the 002 reflection of MgO and Fe are observed for all films, indicating a $c$-axis orientation. For thickness $d\geq90\,\rm nm$, a 110 component is observed whose intensity becomes gradually stronger with increasing $d$. In addition, the ratio of the diffraction intensity for the 110 and 004 is increased with $d$ (Table\,\ref{tab:table1}). This 110 component is also visible in the 103 pole figure measurement, and its epitaxial relation to the substrate is (110)[001]Ba-122$\|$(001)[110]MgO and (110)[001]Ba-122$\|$(001)[$\overline{1}$10]MgO. It should be noted, however, that the amount of the 110 component is small since the ratio of the diffraction intensity for the 110 and 004 is less than 0.01 for all films. Here, the corresponding value for a randomly oriented grain is 3.28 (ICDD card number 01-077-6875).  Indeed, this small amount of 110 component does not compromise the crystalline quality as shown in Table\,\ref{tab:table1}. The full width at half maximum (FWHM), $\Delta\omega$, of the 004 rocking curve and the average $\Delta\phi$ of the 103 reflection of Co-doped Ba-122 are getting smaller with increasing $d$.

\begin{figure}
	\centering
		\includegraphics[width=8.5cm]{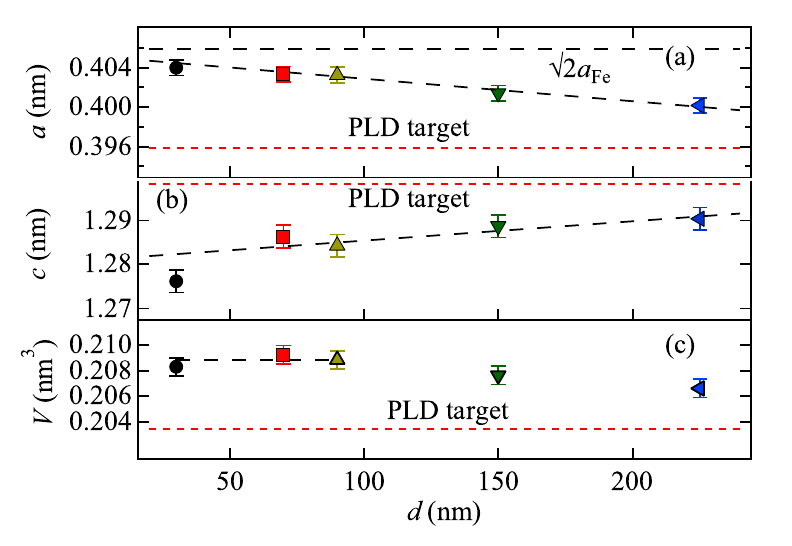}
		\caption{(a) The lattice parameter $a$ is observed to decrease with $d$. (b) Correspondingly, the lattice parameter $c$ are increased with $d$. $a_{\rm Fe}$ is the lattice parameter of Fe. (c) The unit cell volume is almost constant with $d$ in the range of $30\,{\rm nm}\leq d\leq 90\,{\rm nm}$, while the thicker films deviate from this trend. Lines are guide to the eye.} 
\label{fig:figure4}
\end{figure}

Another prominent feature is a shift of the 00$l$ reflections to lower angles with increasing $d$, indicating an increase in the lattice parameter $c$. Calculated lattice parameters $c$ of the films using the Nelson-Riley function are increasing from 1.274\,nm ($d=30\,{\rm nm}$) to 1.289\,nm ($d=225\,{\rm nm}$)\,\cite{09}, as shown in Table\,\ref{tab:table1}. 1. In this calculation, the 002 reflection is omitted in order to avoid excessive extrapolation. The correlation between the layer thickness and both the in-plane and the out-of-plane lattice parameters evaluated by RSM are exhibited in figures\,\ref{fig:figure4}\,(a) and (b). The lattice parameter $a$ is observed to decrease with $d$, while the out-of-plane lattice parameter behaves the opposite way. The evaluated lattice parameter $c$ by both methods (i.e. RSM and the $\theta\rm/2\theta$-\,scans) are almost identical for all the films within the experimental uncertainty, albeit high angle data of $2\theta\geq160\,^\circ$ are not measured in the $\theta\rm/2\theta$-\,scans. Since the FeAs tetrahedron in the Ba-122 bonds coherently to bcc Fe\,\cite{10}, the lattice parameter $a$ of a thin Co-doped Ba-122 layer is close to that of Fe multiplied by $\sqrt2$ (0.4053\,nm), which is slightly larger than $a$ of bulk Co-doped Ba-122 (i.e. PLD target), suggesting tensile strain in the film. The lattice parameter of Fe is almost constant at around 0.287\,nm confirmed by RSM. Here the respective lattice misfit of Fe/MgO and Co-doped Ba-122/Fe are -3.9\% and -2.4\%.
 
\begin{table}
\centering
\caption{\label{tab:table1}Surface roughness, average FWHM values of the $\phi$-scans and the $\omega$-scans, lattice parameter $c$ and diffraction intensity ratio, $I_{\rm 110}$/$I_{\rm 004}$, for Co-doped Ba-122 thin films with different layer thickness, $d$.}
\begin{tabular}{cccccc}
\br
$d$\,(nm)&$R_{\rm rms}$\,(nm)&$\Delta\omega(^\circ)$&$\Delta\phi(^\circ)$&$c$ (nm)&$I_{\rm 110}$/$I_{\rm 004}$\\
\hline
30&1.14&0.82&0.99&1.274&n.\,d.\\
70&0.85&0.96&1.26&1.285&n.\,d.\\
90&1.47& 0.65 & 0.88&1.284&$\leq 0.001$\\
150&0.83& 0.67 & 0.92&1.288&0.003\\
225&2.11& 0.60 & 0.81&1.289&0.007\\
\br
\end{tabular}
\end{table}

\begin{figure}
	\centering
			\includegraphics[width=8.5cm]{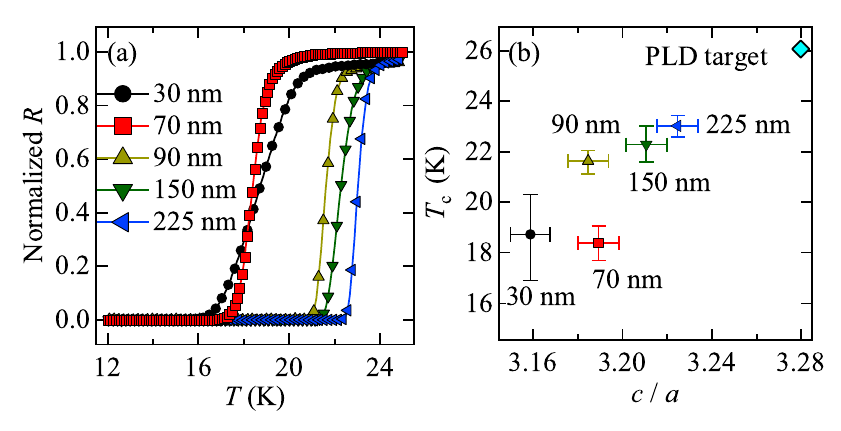}
		\caption{(a) The normalized resistive traces of the Fe/Ba-122 bilayers show a gradual increase in the $T_{\rm c}$ with increasing $d$. (b) The $T_{\rm c}$ of the Co-doped Ba-122 film is sensitive to the lattice distortion. Errors of the $T_{\rm c}$ is defined as a delta $T_{\rm c}$.} 
\label{fig:figure5}
\end{figure}

The unit cell volume, $V=a^2c$, of Co-doped Ba-122 films is almost constant with $d$ in the range of $30\,{\rm nm}\leq d\leq 90\,{\rm nm}$, while the thicker films deviate from this trend presumably due to a relatively large contents of grain boundaries, GBs (figure\,\ref{fig:figure4}(c)). Nevertheless, the volume for all the films is larger than that of the bulk Co-doped Ba-122, which might be due to As deficiency. The correlation between the lattice parameters and As deficiency for Co-doped Ba-122 is not clear. However, both the in-plane and the out-of-plane lattice parameters of As-deficient LaFeAsOF are enlarged compared with that of the stoichiometric sample\,\cite{11}, resulting in a larger lattice volume of the As-deficient sample.
 
Shown in figure\,\ref{fig:figure5}\,(a) are the normalized resistive traces of the Co-doped Ba-122 with different $d$. The film with $d=70\,{\rm nm}$ shows the lowest $T_{\rm c}$ of around 18\,K. The 30\,nm thick film shows a broad transition width, which is a direct consequence of the small grain size together with poor connectivity. Another prominent feature is a clear jump of $T_{\rm c}$ between 70\,nm and 90\,nm thickness. From the $\theta\rm/2\theta$-\,scans in figure\,\ref{fig:figure3}, the 110 component is absent in the 70\,nm thick film whereas the 90\,nm thick film contained a small amount of the 110 grains. For $d \geq 90\,{\rm nm}$, the $T_{\rm c}$s are gradually improved up to 23\,K together with sharpening of the transition width by increasing $d$. It is clear from figure\,\ref{fig:figure5}\,(b) that $T_{\rm c}$ of the films improves with $c/a$, which is consistent with our previous results for films on different substrates\,\cite{06}.

\begin{figure}
	\centering
			\includegraphics[width=8.5cm]{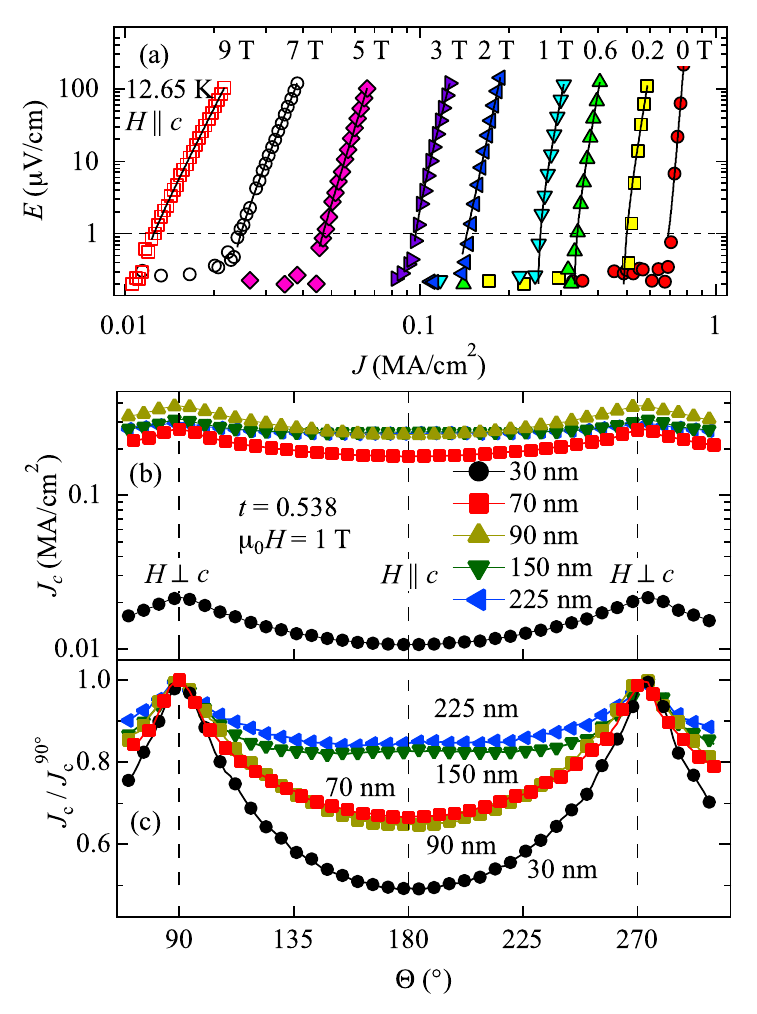}
		\caption{(a) The $E-J$ curves for the Co-doped Ba-122 film with $d=225\,\rm nm$ measured in various magnetic fields at 12.65\,K. Applied magnetic fields are parallel to the $c$-axis of the film. (b) The $J_{\rm c}(\Theta)$ of the films with various $d$ measured in an applied field of 1\,T at a reduced temperature of 0.538. The data of the 225\,nm thick film are not visible, since it shows the almost identical form of $J_{\rm c}(\Theta)$ as the 150\,nm thick film. (c) $J_{\rm c}(\Theta)$ presented fig.\,\ref{fig:figure6}\,(b) are normalized to the value at $\Theta=90^\circ$.} 
\label{fig:figure6}
\end{figure}

The $E-J$ curves for the Co-doped Ba-122 film with $d=225\,\rm nm$ measured in various magnetic fields at 12.65\,K show a power-law relation, indicative of current limitation by depinning of flux lines rather than GB effects (figure\,\ref{fig:figure6}\,(a)). As stated earlier, thicker films contained a small amount of the 110 component. However, the film does not show any sign of weak-link behavior.

The $J_{\rm c}(\Theta)$ measured at a reduced temperature of $t=0.538$ ($t=T/T_{\rm c,0}$, where the $T_{\rm c,0}$ is onset temperature of zero resistance) are exhibited in figure\,\ref{fig:figure6}\,(b). All the films except the 30\,nm thick film can carry a high $J_{\rm c}$ of over $0.18\,{\rm MA/cm^2}$ in the whole angular range. The 30\,nm thick film shows one order of magnitude lower $J_{\rm c}(\Theta)$ than that of the other films, which is due to the small grain size together with poor connectivity. Indeed, $E-J$ curves of this film show a non-ohmic linear differential, NOLD, signature, indicative of $J_{\rm c}$ limitation by GBs\,\cite{12}.

Another prominent feature is a shift upward of $J_{\rm c}$ at $\Theta=180^\circ$ ($H$\,$\parallel$\,${c}$) as the thickness of Co-doped Ba-122 films increases. In particular a small $c$-axis angular peak is observed for $d\geq150\,\rm nm$ (fig.\,\ref{fig:figure6}\,(c)). These films show very similar forms of $J_{\rm c}(\Theta)$ with a low $J_{\rm c}$ anisotropy, $\gamma_{\rm J}=J_{\rm c}(90^\circ)/J_{\rm c}(180^\circ)$. For example, $\gamma_{\rm J}$ of the film with $d\geq150\,\rm nm$ is only around 1.2, whereas the corresponding values of the 90\,nm and 30\,nm thick films are 1.5 and 2, respectively. These results suggest that the GBs may contribute to the pinning along the $c$-axis\,\cite{13}, which opens the opportunity to tune $\gamma_{\rm J}$ by controlling the amount of the 110 textured component.

The implementation of very thin Fe buffer layers also yields the 110 textured component\,\cite{05}. However, the crystalline quality and the superconducting properties of Co-doped Ba-122 are compromised due to the presence of too many GBs. Hence, there may exist a threshold of the amount of GBs at which deterioration of the structural and superconducting properties sets in.

\section{Conclusion}
The effect of the superconducting layer thickness on the structural and superconducting properties of Co-doped Ba-122 films has been investigated. The texture quality and the superconducting transition temperature are improved by increasing the layer thickness due to stress relief. Increasing the layer thickness yields an additional 110 textured component, creating GBs. However, these GBs may constitute $c$-axis  pinning centers within a certain amount, which leads to an increase in $J_{\rm c}$ at $H$\,$\parallel$\,${c}$ without compromising the structural and superconducting properties.

\ack
The authors thank J.\,Scheiter for help with FIB cut samples and E.\,Barbara for help with the AFM observation. We are also grateful to M.\,K\"{u}hnel and U.\,Besold for their technical support and S.\,F\"{a}hler for his RHEED software.

\end{document}